\documentstyle[aps,preprint,epsf,epsfig]{revtex}
\newcommand{\beq}{\begin{equation}}
\newcommand{\eeq}{\end{equation}}
\newcommand{\bea}{\begin{eqnarray}}
\newcommand{\eea}{\end{eqnarray}}
\newcommand{\ba}{\begin{array}}
\newcommand{\ea}{\end{array}}
\newcommand{\bc}{\begin{center}}
\newcommand{\ec}{\end{center}}
\newcommand{\lsimeq}{\stackrel{<}{\scriptstyle\sim}}

\newcommand{\bml}{\begin{mathletters}}
\newcommand{\eml}{\end{mathletters}}
\begin{document}
\draft
\preprint{}
\title{ Bose-Stimulated Raman Adiabatic Passage in Photoassociation}
\author{Matt Mackie, Ryan Kowalski, and Juha Javanainen}
\address{Department of Physics, University of Connecticut, Storrs, CT
06269-3046}
\date{\today}
\maketitle

\begin{abstract}
We analyze coherent two-color photoassociation of a Bose-Einstein
condensate, focusing on stimulated Raman adiabatic passage~\mbox{(STIRAP)}
in free-bound-bound transitions from atoms to molecules. This problem is
of particular interest since STIRAP is predicted to be necessarily absent
in the nondegenerate case [Javanainen and Mackie, Phys.\ Rev.\ A {\bf 58},
R789 (1998)]. However, Bose-stimulation enhances the free-bound dipole
matrix element for an atomic condensate, and photoassociative STIRAP turns
out to be a viable mechanism for converting an atomic condensate to a
molecular condensate with near-unit efficiency.
\end{abstract}


\narrowtext

As the field of Bose-Einstein condensation~(BEC) in dilute gases continues
to develop, candidate systems for BEC are being discovered on a
regular basis. Besides the original alkalis~\cite{BEC}, there is the
recent observation~\cite{HBEC} of BEC in hydrogen~\cite{HBEC76}.
Meanwhile, a spin-polarized helium condensate is not out of the
question~\cite{HeBEC}, and recent theoretical work has shown that coherent
photoassociation might be used to produce a degenerate molecular
gas~(MBEC) from an already-Bose-condensed sample of
atoms~\cite{DRUM98,JJMM99}. Here we consider the formation of MBEC
using coherent photoassociation.

Photoassociation~(PA) is the process in which a pair of atoms interacts
with a photon, thereby making a transition from the two-atom continuum to a
bound state of the molecule. Quantization of the molecular dissociation
continuum allows one to describe such free-bound transitions
using the standard techniques of few-level quantum
optics~\cite{JJMM98,MMJJ99}. Beyond providing
a few-level framework, such a quasicontinuum approach has also led to a
matter-quantized formulation of photoassociation analogous to the analysis
of  second-harmonic generation of light~\cite{DRUM2,JJMM99}. In this
context of nonlinear matter optics~\cite{LENZ}, atom-molecule conversion
displays coherent BEC-MBEC oscillations, adiabatic following, and
non-classical collapse and revivals~\cite{JJMM99,DISS}.

Nonetheless, one-color free-bound photoassociation generally occurs to an
excited electronic state of the molecule, and the subsequent irreversible
losses, whether due to photodissociation~(PD) or spontaneous decay, tend
to negate the benefits of the coherence. In this Letter we develop coherent
two-color free-bound-bound photoassociation, where the primary
photoassociated molecules are transferred with another laser field to a
stable molecular state. The goal is to state-selectively convert a BEC
into an MBEC, while averting irreversible losses from electronically
excited molecules. We consider pulsed free-bound and bound-bound
couplings that occur in the counter-intuitive order~\cite{STIRAP}, and so
correspond to stimulated Raman adiabatic passage~(STIRAP) from atoms to
molecules. Beyond its distinctly nonlinear character, this problem is
of particular interest since we have previously argued for the {\it
absence} of free-bound-bound
\mbox{STIRAP}~\cite{JJMM98}. However, the present work illustrates that
such transitions are possible in the case of a condensate, since
the free-bound dipole matrix element is Bose-enhanced due to the fact that
many atoms are in the same quantum state. A similar statistical effect
may also increase hot pion production in nuclear physics~\cite{PIONS}, and
speed up the decay rate of the inflation field in cosmology~\cite{COSMO}.

The development herein is outlined as follows. First, we review the
many-body enhancement for the PA dipole matrix element of a
condensate, and discuss the implications for coherent free-bound-bound
STIRAP. Next we take a semi-classical approach to identify the nonlinear
counterpart of the dark state~\cite{DARKSTS} that
contains no primary photoassociated molecules. Given that Bose-enhancement
allows a counter-intuitive pulse scheme, the limit
$t\rightarrow -\infty$ finds the dark state with all atoms,
while in the limit $t\rightarrow +\infty$ the dark state contains only
molecules. It then appears possible that free-bound-bound STIRAP of an
atomic condensate will produce a similarly degenerate gas of molecules.
The remaining work establishes conditions that allow such
adiabatic following to occur.

\begin{figure}
\centering
\epsfig{file=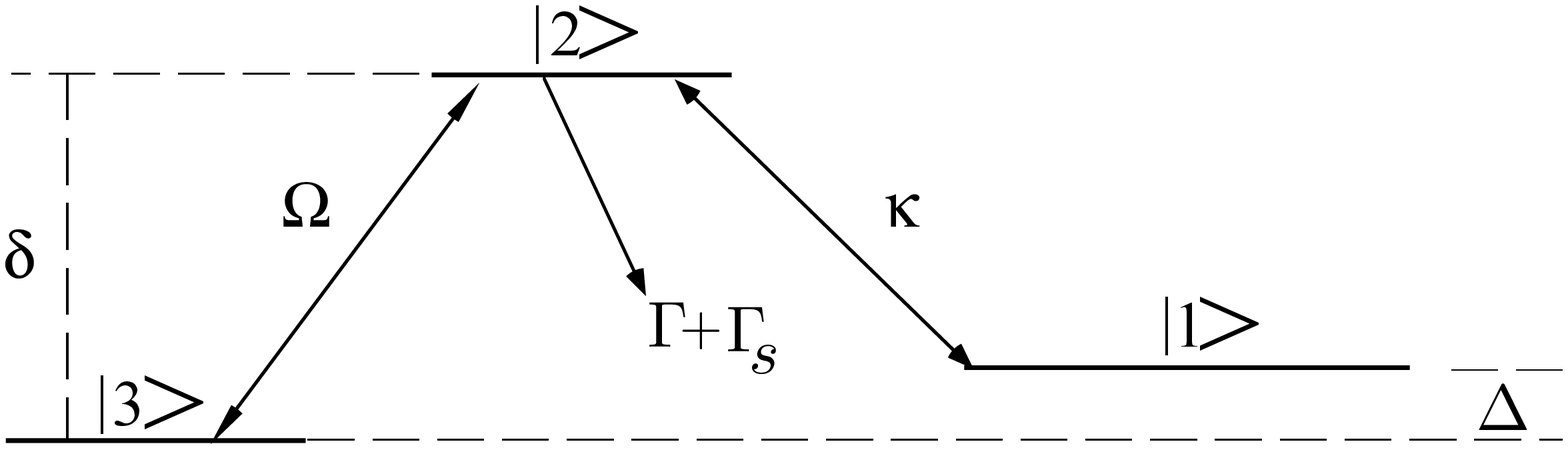,width=13cm,height=10cm}
\caption{Three-level illustration of coherent free-bound-bound
photoassociation, where $N$
atoms have assumedly Bose-condensed into state $|1\rangle$.
The free-bound and bound-bound  Rabi frequencies are $\kappa$ and $\Omega$,
respectively. Similarly,
the two-photon and intermediate detunings are $\Delta$ and $\delta$.
Irreversible losses are due to either
photodissociation, at the rate $\Gamma$, or spontaneous decay, at the rate
$\Gamma_S$. The difference
from the familiar three-level scheme is that the free-bound interaction
involves three particles, and is
therefore nonlinear.}
\label{THREELS}
\end{figure}

Turning to the situation of Fig.~\ref{THREELS}, we assume that $N$
identical atoms of mass $m$ have condensed into the same one-particle
state, say, a plane wave state with wave vector ${\bf k}=0$.
Photoassociation  removes two atoms from this
state $|1\rangle$ and creates an excited molecule in state $|2\rangle$,
with the translational energy due to the recoil momentum of the free-bound
photon equal to $\hbar^2{\bf q}_1^2/4m$. Including a second laser
frequency, bound-bound transitions remove excited molecules from state
$|2\rangle$ and create stable molecules in state $|3\rangle$. The
translational energy per stable molecule arises from the combined recoil
momenta of two photons, and equals $\hbar^2({\bf q}_1-{\bf q}_2)^2/4m$. In
second-quantized notation, we denote the Boson annihilation operators for
atoms, primarily photoassociated molecules, and stable molecules,
respectively, by $a$, $b$, and $g$.

The laser-matter interactions that drive the atom-molecule and
molecule-molecule transitions are written in terms of their
respective Rabi frequencies,
$\kappa=d_1 E_1/ 2\hbar$ and
$\Omega=d_2 E_2 /
2\hbar\,$. Here the amplitude of the electric field driving a given
transition is $E_i\,$, and $d_i$ is the
corresponding dipole matrix element~($i=1,2$). Lastly, we define the
two-photon and intermediate detunings, $\Delta$ and $\delta$, both with
due inclusion of photon recoil energies. Analogously to
Ref.~\cite{JJMM99}, the  Hamiltonian for the system is
\beq
{H\over\hbar} =\hbox{$1\over2$}\Delta a^\dagger a +\delta b^\dagger b
   -\hbox{$1\over2$}\kappa
\left( aab^\dagger + a^\dagger a^\dagger b\right)
   -\Omega\left(bg^\dagger + b^\dagger g\right).
\eeq

The Bose-enhancement of the free-bound dipole matrix element demonstrated as
follows. First we consider the Heisenberg equations of motion, which determine the
time evolution of the system according to
\bml
\bea
\dot{a}&=&-i\left({1\over 2}\Delta a - \kappa a^\dagger b\right), \\
\dot{b}&=&-i\left(\delta b - {1\over 2}\kappa aa - \Omega g \right), \\
\dot{g}&=&i\Omega b\,.
\eea
\label{EOM}
\eml
Now, since the number of particles is conserved,
\beq
a^\dagger a + 2(b^\dagger b + g^\dagger g)=N,
\label{COM}
\eeq
it is clear that $a,b,g \sim \sqrt{N}$. Hence, we define scaled boson
operators of order unity as $x\rightarrow x'=x/\sqrt{N}$, with
$x=a,b,g$. Dropping the primes, the conserved
quantity~(\ref{COM}) is normalized to unity, and the equations of motion
are given by
\bml
\bea
\dot{a}&=&-i\left({1\over 2}\Delta a - \chi a^\dagger b\right), \\
\dot{b}&=&-i\left(\delta b - {1\over 2}\chi aa - \Omega g \right), \\
\dot{g}&=&i\Omega b\,.
\eea
\label{SEOM}
\eml
 From Eqs.~(\ref{SEOM}), the many-body Bose-enhancement of the free-bound
dipole matrix element is
evident in the scaled Rabi frequency $\chi=\sqrt{N}\kappa$.

As it happens, the present STIRAP analysis depends crucially on the fact
that the bare free-bound coupling $\kappa$ is scaled by the factor
$\sqrt{N}$, while that for the bound-bound transition, $\Omega$, is
unchanged. To see why, we recall photoassociation in terms of our
quasicontinuum model~\cite{JJMM98,MMJJ99}, which represents the
dissociation continuum of the molecule as an infinite number of discrete
energy levels separated equally by $\hbar\epsilon$. The correct free-atom
results are recovered by taking the limit $\epsilon\rightarrow 0$. But,
the bare free-bound coupling is given in terms of the frequency spacing as
$\kappa\sim\sqrt{\epsilon}$, or in terms of the corresponding quantization
volume as
$\kappa\sim1/\sqrt{V}$. In the free-atom limit ($\epsilon\rightarrow 0$ or
$V\rightarrow\infty$) we then find that
$\kappa\rightarrow 0$.  While this observation does not condemn
photoassociation of a nondegenerate gas in the thermodynamic
limit~\cite{JJMM98,MMJJ99}, $V\rightarrow\infty$ and $N\rightarrow\infty$ with
$\rho=N/V$ constant, it does imply that $\kappa(t)\ll\Omega(t)$~for~all~$t$. Hence,
a counter-intuitive reversal of the coupling strengths of the
pulses~\cite{STIRAP}, $\Omega(t)\gg\kappa(t)$ for $t\rightarrow -\infty$ and
$\kappa(t)\gg\Omega(t)$ for $t\rightarrow +\infty$, cannot be achieved.
It is therefore the {\it absence} of STIRAP that has been predicted for
nondegenerate free-bound-bound transitions~\cite{JJMM98}. However, as
shown above, in a condensate the bare Rabi frequency of a nondegenerate
gas $\kappa$ is scaled by the Bose enhancement factor $\sqrt{N}$, which
leads to a finite value even in the thermodynamic limit~\cite{JJMM99};
$\chi\sim\sqrt{N/V}\sim\sqrt{\rho}$. This observation will open the door to using
STIRAP as a means to create a stable molecular condensate.

In order to facilitate an analytical solution, we define the ``Kamiltonian"
for the system by adding a
multiple of the conserved particle number to the Hamiltonian
\beq
K=H-\hbar\mu\left[a^\dagger a+2\left(b^\dagger b+g^\dagger g\right)\right],
\eeq
where the real constant $\hbar\mu$ is identified as the chemical potential
per atom. The
Heisenberg equations of motion for the unit-scaled operators become
\bml
\bea
\dot{a}&=&i\left[\left(\mu-{1\over 2}\Delta\right) a + \chi a^\dagger
b\right], \\
\dot{b}&=&i\left[\left(2\mu-\delta\right) b + {1\over 2}\chi aa +
\Omega g \right], \\
\dot{g}&=&i\left(2\mu g+\Omega b\right).
\eea
\label{BEQM}
\eml
 From this point onward we also resort to the semi-classical approach analogous
to the Gross-Pitaevskii approximation used to describe an alkali
condensate. Accordingly, the quantities $a$, $b$ and $g$ are now
$c$-numbers, rather than operators.

We are looking for adiabatic solutions
to the time-evolution equations~(\ref{BEQM}) for transient couplings
$\chi(t)$ and $\Omega(t)$. Denoting the characteristic Rabi frequency
scale for the light pulses by $R$ and the characteristic pulse width by
$T$, the adiabatic approximation ($\dot{x}\approx 0,x=a,b,g$) should be
valid when the evolution time scale of the system is short compared to the
time scale of the pulses; for instance when $RT\gg 1$. We return to the
adiabatic condition in a moment, and for the time being simply assume
time scales for the problem such that $\dot{x}\approx 0$ is valid.

Besides the steady state, we now specify exact two-photon resonance
($\Delta=0$). To economize the ensuing expressions, we also choose the
Bose-enhanced Rabi coupling $\chi$ as the scale for our frequencies by writing
$\Omega=\bar\Omega\chi$, $\delta=\bar\delta\chi$. Now, we have already assumed
the Rabi frequencies to be real. Without sacrificing generality, this will allow us to
consider the amplitudes $a$, $b$, and $g$ as strictly real. Additionally, for
counter-intuitive pulses the limits $t\rightarrow (-\infty,+\infty)$ correspond to the
limits ${\bar\Omega}\rightarrow (\infty,0)$. Thus we neglect any
solution which is not similarly real for {\it all} values of $\bar{\Omega}$ from
$0$ to $\infty$. Discarding
also solutions differing only by redundant signs, there simply remains
\bml
\bea
\mu_0&=&0,\\
a_0&=&\sqrt{\bar\Omega\left(\sqrt{2+\bar\Omega^2}-\bar\Omega\right)}\,,\\
b_0&=&0,\\
g_0&=&-{1\over 2}\,\left(\sqrt{2+\bar\Omega^2}-\bar\Omega\right);
\eea
\label{DARKST}
\eml
\bml
\bea
\mu_\pm&=&\pm\,{\bar\Omega^2
+\bar{\delta}\left(\bar{\delta}\pm\sqrt{\bar\Omega^2+\bar{\delta}^2}\right)
\over 2\sqrt{\bar\Omega^2+\bar{\delta}^2}}\,\chi, \\
a_\pm&=&0, \\
b_\pm&=&{1\over
2}\,\sqrt{1\pm{\bar{\delta}\over\sqrt{\bar\chi^2+\bar{\delta}^2}}}\,,\\
g_\pm&=&-{1\over
2}\sqrt{1\pm{\bar{\delta}\over\sqrt{\bar\Omega^2+\bar{\delta}^2}}}\nonumber\\
&&\hspace{1cm}\times\left[{\bar\Omega^2
+\bar{\delta}\left(-\bar{\delta}\pm\sqrt{\bar\Omega^2+\bar{\delta}^2}\right)
\over \bar\Omega\sqrt{\bar{\Omega}^2+\bar{\delta}^2}}\right].
\eea
\label{NONADSTS}
\eml

It is easy to see that the above results are the nonlinear counterparts
of those obtained from the standard STIRAP analysis~\cite{STIRAP}. The
solution $(\ref{DARKST}$) is the dark state. For a counter-intuitive
pulse sequence, $\bar{\Omega}\gg 1$ for $t\rightarrow-\infty$ and
$\bar{\Omega}\ll 1$ for $t\rightarrow+\infty$, the dark state~(\ref{DARKST})
initially consists of atoms ($a_0\rightarrow 1,\,g_0\rightarrow 0$), while
the final dark state is all molecules~($a_0\rightarrow
0,\,g_0\rightarrow 1/\sqrt{2}$). At no intervening time is there any
population in the intermediate molecular state ($b_0\equiv0$). If the laser
pulses allow for adiabatic evolution, an atomic condensate is converted to
a molecular condensate without any loss from the intermediate state.

\begin{figure}
\centering
\epsfig{file=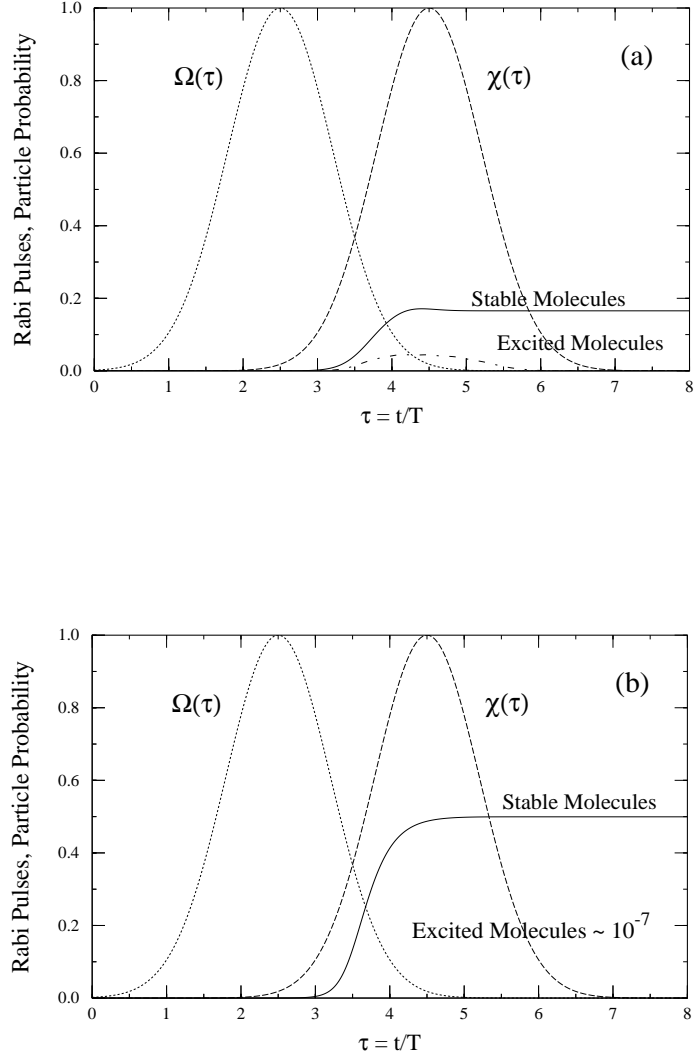,width=13cm,height=17cm}
\caption{Transient free-bound-bound photoassociation of a BEC for a
counterintuitive pulse order. The pulses are of equal height,
$\chi_0=\Omega_0=R$, so that $R=1$ sets the unit of frequency. The
intermediate detuning is $\delta=1$, and the pulse delays are $D_1=4.5\,T$
and $D_2=2.5\,T$. Also, recall that $N$ atoms gives a maximum of $N/2$
molecules, so that $|g|^2=1/2$ represents complete conversion of the
initial condensate. (a) $T=10$ results in a pulse area insufficient to
provide adiabaticity.  (b) As expected, at $T=10^4$ the
initial BEC  is converted
entirely into MBEC.}
\label{NUMSOLNS}
\end{figure}

We have checked these results with an exact numerical solution to the
equations of motion~(\ref{BEQM}), examining the intuitive adiabatic
condition $RT\gg 1$ as well. In Fig.~\ref{NUMSOLNS}, Gaussian pulses of
the form $\chi(t)= R\exp\left[-(t-D_1)^2/T^2\right]$ and
$\Omega(t)=R\exp\left[-(t-D_2)^2/T^2\right]$ illustrate that an
insufficient pulse area ($RT$) results in a significantly populated
intermediate state. On the other hand, a pulse area much greater than
unity readily decouples the nonadiabatic states~(\ref{NONADSTS}), allowing
the system to adiabatically follow the dark state as it moves from the
initial BEC to MBEC. In particular, Fig.~\ref{NUMSOLNS}(b) gives the
probability of creating excited molecules as $|b|^2\sim 10^{-7}$, so that
irreversible losses from either photodissociation or spontaneous decay
should be negligible. Coherent free-bound-bound STIRAP is thereby
confirmed.

To improve upon our discussion of the adiabatic approximation, we focus
specifically on the effect of pulse overlap. We apply the textbook
criterion\cite{SCHIFF} for adiabaticity to the eigenvector
$\psi=(a,b,g)^T$. From Eqs.~(\ref{DARKST})~and~(\ref{NONADSTS}), the
coupling between the nonadiabatic states, $\psi_\pm$, and the rate of
change of the adiabatic state,
$\dot{\psi}_0$, must therefore be much less than the spacing between the
respective chemical potentials,
$\left| \psi_\pm^T\,\dot{\psi}_0 \right| \ll \left|\mu_0-\mu_\pm\right|$.
Furthermore, we restrict our analysis to zero intermediate
detuning~($\bar{\delta}=0$), and introduce the arbitrary pulse shapes
$\chi(t)=R f_1(\tau)$ and $\Omega(t)=R f_2(\tau)$, where $\tau=t/T$ is a
dimensionless time. The adiabatic condition is then $F\ll RT$,
where the (dimensionless) nonlinear adiabatic factor, $F$, is given as
\bea
F&=&{|\dot{f}_1(\tau)/f_1(\tau) - \dot{f}_2(\tau)/f_2(\tau)|
   \over 2\,\sqrt{2f_1^2(\tau)+f_2^2(\tau)}} \nonumber\\
&&\hspace{1cm}\times \left|
f_2(\tau)/f_1(\tau)-\sqrt{2+f_2^2(\tau)/f_1^2(\tau)}\, \right|\,.
\label{NLADF}
\eea

\begin{figure}
\centering
\epsfig{file=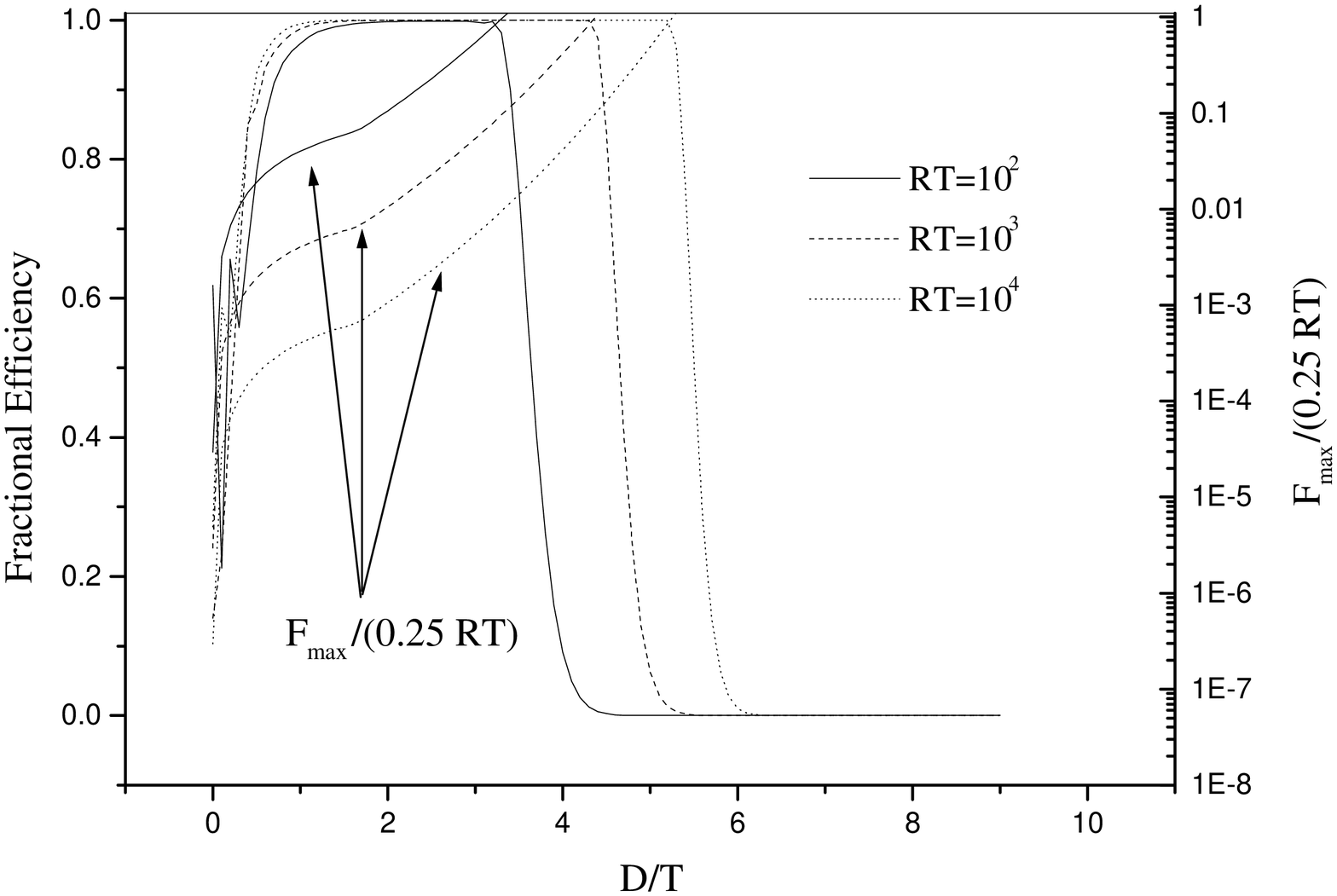,width=13cm,height=10cm}
\vspace{1cm}
\caption{Efficiency of free-bound-bound STIRAP and the nonlinear adiabatic
factor $F$ plotted as a
function of the pulse separation $D=D_1-D_2$. In the region of pulse
overlap, $\tau\approx
(D_1+D_2)/2T$, we determine the maximum value of $F(\tau)$,
$F_{\text{max}}$, and compare it to the numerical results for the fractional
efficiency, $2|g|^2$. For
$R=1$ and $T=(10^2,10^3,10^4)$, it is clear that $F_{\text{max}}/(0.25\,RT)=1$
exactly marks the
breakdown of adiabaticity.}
\label{ADCONTEST}
\end{figure}

We find a dependence on pulse overlap that is qualitatively similar to
the results in Ref.~\cite{STIRAP} for ordinary STIRAP. In particular, if the two
pulses vanish concurrently~(such that their ratio is finite), the fraction in
Eq.~(\ref{NLADF}) diverges, and the adiabatic condition, $F\ll RT$, is
violated. Additionally, an increase in pulse area $RT$ may provide
adiabaticity despite a poor pulse overlap. Using Gaussian pulse shapes,
these observations may be quantified as follows. In the vicinity of pulse
overlap, at dimensionless times \mbox{$\tau\simeq (D_1+D_2)/2T$}, we
determine the maximum value of
$F(\tau)$, $F_{\text{max}}$, as a function of the delay between the pulses,
$D=D_1-D_2$. Numerically,
breakdown of adiabaticity occurs at a specific pulse separation when the
fractional efficiency,
$2|g|^2$, is no longer of order unity; hence, the value of $F_{\text{max}}$ at this
point defines the adiabatic condition in terms of $D$. The results for
$RT=(10^2,10^3,10^4)$ are shown in
Fig.~\ref{ADCONTEST}. We find that the adiabatic approximation is
valid for pulse overlaps
satisfying $F_{\text{max}}(D)\lsimeq 0.25\,RT$. Incidentally, we find exactly
the same result for
peaked-exponential pulse shapes, $f_{1,2}(\tau)=\exp(-|t-D_{1,2}|/T)$.

It remains to discuss a few items that we have so far ignored. The first regards
the explicit role of photodissociation~(PD). Consistent with the
quasicontinuum model~\cite{JJMM98,MMJJ99,DISS}, we consider PD as
irreversible decay from the intermediate state to the atomic continuum. Hence, if
there is no intermediate-state population, there is no photodissociation. Second,
our results are of course valid regardless of trapping of the atoms and molecules,
provided that the time scale for coherent free-bound-bound STIRAP is shorter than
the time scale for the motion of the atoms and/or molecules in the trap. Third, if laser
intensities permit STIRAP during a time much shorter than the time scales
for collisions between atoms and molecules, collisions are negligible as
well. Fourth, we have not mentioned the divergence of the adiabatic factor
in the wings of the pulses~\cite{STIRAP}. This is an artifact due to the
mathematical shape of the pulse, and can be avoided if, in the region of
interest, the shapes satisfy $d(\log\bar{\Omega})/d\tau=0$. In fact, this is
the case for the peaked exponentials mentioned above.

In conclusion, we hold the line on the absence of \mbox{STIRAP} in
non-degenerate free-bound-bound transitions, while at the same time
proposing the counter-intuitive pulse scheme as a possible mechanism for
creating a molecular condensate from an initial BEC. This dichotomy arises
because, for the case of a condensate, all $N$ atoms are in the same
quantum state, and the subsequent Bose-enhancement of the free-bound
dipole matrix element enables a counter-intuitive reversal of the Rabi
frequencies associated with the light pulses. Our numerical trials have
confirmed that STIRAP should take place in two-color photoassociation, and
validated a simple quantitative criterion for adiabatic atom-molecule
conversion.

This work was supported in part by the NSF, Grant No. PHY-9801888, and the
Research Experience for Undergraduates Program. Additional support provided
by NASA, Grant No. NAG8-1428 and the Connecticut Space Grant College
Consortium.



\begin {thebibliography}{99}
\bibitem{DRUM2} The SHG analogy is also drawn in Ref.~\cite{DRUM98},
independently of the quasicontinuum model.
\end{thebibliography}
\end{document}